\documentclass[twoside,11pt]{article}
\usepackage{a4wide,pictex,cite,latexsym,amsfonts,amssymb,exscale,epsfig,bbm}
\usepackage[centertags,sumlimits,intlimits,namelimits,reqno]{amsmath}

\usepackage{amsthm}
\theoremstyle{definition}
\newtheorem{theorem}{Theorem}[section]

\newtheorem{proposition}[theorem]{Proposition}
\newtheorem{remark}[theorem]{Remark}
\newtheorem{definition}[theorem]{Definition}

\makeatletter
\renewcommand{\theequation}{\thesection.\arabic{equation}}%
\@addtoreset{equation}{section}

\newcounter{mathletter}%
\newcommand{\bmathletter}{%
  \refstepcounter{equation}%
  \setcounter{mathletter}{\value{equation}}%
  \setcounter{equation}{0}%
  \@ifundefined{chapter}{%
    \renewcommand{\theequation}{%
      \mbox{\thesection.\arabic{mathletter}\alph{equation}}}}{%
    \renewcommand{\theequation}{%
      \mbox{\thechapter.\arabic{mathletter}\alph{equation}}}}}%
\newcommand{\emathletter}{\setcounter{equation}{\value{mathletter}}}%
\newenvironment{mathletters}{\bmathletter}{\emathletter}%
\makeatother

\newenvironment{myenumerate}{%
  \begin{enumerate}
  \setlength{\partopsep}{0pt}
  \setlength{\parskip}{0pt}}{\end{enumerate}}

\def\openone{\mathbbm{1}}%
\def\pacs#1{\noindent PACS: #1\par}%
\def\keywords#1{\noindent key words: #1\par}%
\def\acknowledgements{\section*{Acknowledgements}}%

\def\dopreprint{\hfill{\small\thepreprint}\\}%
\def\preprint#1{\def\thepreprint{#1}}%
\def\thepreprint#1{}%
\def\sym#1{{\mathcal #1}}
\def\emph#1{{\sl #1\/}}

\newcommand{\ontop}[2]{\genfrac{}{}{0pt}{2}{\scriptstyle #1}{\scriptstyle #2}}

\let\phi=\varphi
\let\theta=\vartheta
\let\epsilon=\varepsilon

\let\hat=\widehat
\let\tilde=\widetilde

\def\dim{\mathop{\rm dim}\nolimits}
\def\End{\mathop{\rm End}\nolimits}
\def\Hom{\mathop{\rm Hom}\nolimits}
\def\Aut{\mathop{\rm Aut}\nolimits}

\def\address#1{\date{{\sl #1}\\\ \\\theversion}\gdef\date##1{}}%
\ifx\@draft\@isdraft
  \def\version#1{\gdef\theversion{(Draft of #1)}}%
\else
  \def\version#1{\gdef\theversion{#1}}%
\fi

\def\mycaption#1#2{%
  \begin{quote}
  \caption{\label{#1}#2}
  \end{quote}}

\def\ie{{\sl i.e.\/}}
\def\eg{{\sl e.g.\/}}

\def\etc{{\sl etc.\/}}
\def\cf{{\sl cf.\/}}

\def\nn{\notag}
\def\eqref#1{(\ref{#1})}%

\def\C{{\mathbbm C}}
\def\N{{\mathbbm N}}
\def\R{{\mathbbm R}}
\def\Z{{\mathbbm Z}}

\def\Calg{C_{\rm alg}}%
\def\Rep{\tilde{\sym{R}}}%
\def\Irrep{\sym{R}}%
\def\Wloop{W(\{\tau_{j\kappa}\},\{Q^{(j)}\})}
 
\version{15 January 2001}%
\preprint{DAMTP-2000-84}%

\begin{document}
%
% ==============================================================================

\title{\dopreprint The dual of pure non-Abelian lattice gauge theory\\
       as a spin foam model}
\author{Robert Oeckl\thanks{e-mail: R.Oeckl@damtp.cam.ac.uk}\ \ 
        and Hendryk Pfeiffer\thanks{e-mail: H.Pfeiffer@damtp.cam.ac.uk}}
\address{Department of Applied Mathematics and Theoretical Physics,\\
         Centre for Mathematical Sciences, Cambridge CB3 0WA, UK}
\date{\version}
\maketitle

% ==============================================================================
%
\begin{abstract}
%
% ==============================================================================
  
  We derive an exact duality transformation for pure non-Abelian gauge
  theory regularized on a lattice. The duality transformation can be
  applied to gauge theory with an arbitrary compact Lie group $G$ as
  the gauge group and on Euclidean space-time lattices of dimension
  $d\geq 2$. It maps the partition function as well as the expectation
  values of generalized non-Abelian Wilson loops (spin networks) to
  expressions involving only finite-dimensional unitary
  representations, intertwiners and characters of $G$. In particular,
  all group integrations are explicitly performed. The transformation
  maps the strong coupling regime of non-Abelian gauge theory to the
  weak coupling regime of the dual model. This dual model is a system
  in statistical mechanics whose configurations are spin foams on the
  lattice.
\end{abstract}


% ==============================================================================
%
\section{Introduction}
%
% ==============================================================================

Besides the electric-magnetic duality of the vacuum Maxwell equations, the
first example of a duality transformation relating the strong coupling regime
of one field theoretic system with the weak coupling regime of the same or
another system was probably the Kramers-Wannier transformation for the Ising
model \cite{KrWa41a}. This transformation was generalized to a wide class of
Abelian lattice models (spin models, gauge theories and their higher rank
tensor generalizations) in any Euclidean space-time dimension. For a review
see \eg~\cite{Sa80}. 

In this paper, we generalize the transformation to the case of lattice gauge
theory with a non-Abelian gauge group $G$ (our proofs are valid for compact
Lie groups and finite groups). The resulting dual model generalizes the
well-known results for the Abelian case and is described in terms of the
finite-dimensional unitary representations of $G$, their representation
morphisms (intertwiners) and the character expansion of the Boltzmann weight.
In particular, all group integrations are explicitly performed.

The method we use is the Peter-Weyl decomposition of the algebra
$\Calg(G)$ of representation functions of $G$, the `algebraic
functions' on $G$. This decomposition can be viewed as a
generalization of Fourier transformation to functions on a compact
non-Abelian Lie group. It is convenient to exploit the Hopf algebra
structure of $\Calg(G)$ and to employ a purely algebraic description
of the Haar measure. The duality transformation follows the lines of
the well-known Abelian case, but some attention and geometric
intuition is necessary to make the generalized gauge constraint appear
in a local form in the dual model.

The duality transformation establishes the equality of the partition
function and the expectation value of the non-Abelian Wilson loop
(the generic gauge invariant expression which is
given by a spin network) with their corresponding purely algebraic
expressions in the dual model. The dual model is found to have a
Boltzmann weight of such a form that the strong coupling regime of
non-Abelian lattice gauge theory is mapped to the weak coupling regime
of the dual model. In addition, the strong coupling expansion can be
applied to this reformulation of non-Abelian gauge theory in a
systematic way. 

The duality relation is stated in Theorems~\ref{thm_dualpartition1}
and~\ref{thm_dualwilson1} which form the main result of this paper.
The dual model is a system in statistical mechanics whose
configurations are spin foams on the lattice. These configurations are
assigned Boltzmann weights and are subject to certain
constraints. Spin foams have been introduced in the study of quantum
gravity, see \eg~\cite{Re94,Iw95,ReRo97,Ba98} and the recent
introductory article~\cite{Ba99}. The configurations of our dual model
are closed spin foams on the lattice according to the definition given
in~\cite{Ba98}. The transformation thus provides an explicit example
for the relation of lattice gauge theory with a particular spin foam
model.

The dual model reduces to the known results for non-Abelian gauge
theory in $2$ dimensions where the partition function is particularly
simple, as well as to the known results for Abelian lattice gauge
theory in arbitrary dimension, \ie\ in the cases $G=U(1)$, $\Z$ and
$\Z_n$.

Whereas the duality transformation in the Abelian case is known to
work in a similar way for spin models, gauge theories and higher rank
tensor models on the lattice~\cite{Sa80}, this is not the case for the
non-Abelian generalization. Even though it can be applied to spin
models in a straight forward way, the resulting `gauge' constraint
cannot be cast in a local form.

The motivation for deriving a dual description of non-Abelian lattice
gauge theory arises from conceptual issues such as the `dual
superconductor' picture of confinement and from rigorous studies in
the framework of constructive quantum field theory as well as from
technical and numerical problems in lattice gauge theory.

At present only a few ways of studying gauge theories in their strong
coupling regime are known --- in particular if there are no additional
symmetries like supersymmetry. In the famous paper by
Wilson~\cite{Wi74}, the lattice formulation of $U(1)$ gauge theory was
used in conjunction with the high temperature expansion which is known
from statistical mechanics and which plays the role of a strong
coupling expansion.

The duality transformation for Abelian lattice gauge theories (see
\eg~\cite{Sa80}) can be seen on the one hand as a result of attempts
to make this strong coupling expansion systematic. On the other hand
there is the picture of `dual superconductivity' as an explanation of
confinement going back to ideas of t'Hooft and Mandelstam, see
\eg~\cite{BaMy77,Pe78}. Whereas in a superconductor electrically
charged quasi-particles condense and force the magnetic flux into
quantized tubes, the picture is that in a gauge theory with
confinement, magnetic monopoles condense. This leads to the formation
of electric flux tubes which are responsible for the linearity of the
static potential between opposite external electric charges.

In this picture the magnetic monopoles appear as collective
excitations of the gauge theory. They are found to be quantized
topological defects~\cite{FrMa86,FrMa87}. The Abelian duality
transformation allows one to spot these topological degrees of
freedom. All group integrations are performed, and the partition
function of Abelian lattice gauge theory is rewritten in new variables
such that the topological degrees of freedom have the form of
ordinary expectation values of the dual fields.

In the Abelian case, the lattice approach to gauge theories in
conjunction with the duality transformation and related techniques has
lead to a number of remarkable results. We mention the existence of a
phase transition in $U(1)$ lattice gauge theory in $d=4$
\cite{Gu80,FrSp82}, the existence of world-lines of magnetic monopoles
in the same model which behave like infra-particles
\cite{FrMa86,FrMa87}, the fact that these are responsible for the
phase transition by condensation of magnetic monopoles, and finally the
absence of a deconfinement phase transition in $d=3$
\cite{GoMa82}. The picture of dual superconductivity leading to
confinement in the Abelian case is well established, see \eg\ the
study of the monopole degrees of freedom in Monte-Carlo
simulations~\cite{PoWi91} and their properties at the deconfining
phase transition~\cite{JeNe99} of $U(1)$ lattice gauge theory in
$d=4$.

Although there are strong conjectures both from lattice studies of QCD and
from results in supersymmetric Yang-Mills theory that confinement in
non-Abelian gauge theories can be described in a similar way by dual
superconductivity, no analogous approach is available for non-Abelian gauge
theory. The exact duality transformation presented in this paper is meant to
be a first step in this direction.

Finally, we comment on the relations of this work with other
approaches.  Firstly, in the Abelian case, the dual model is again a
gauge theory if certain cohomologies of the space-time lattice are
trivial. The case of general topology is studied
in~\cite{Ja99}. Furthermore, for $SU(2)$ lattice gauge theory, there
are results by explicit computations which relate this theory to
certain simplex models of gravity~\cite{AnCh93,DiPe99}.

Furthermore, there has been an interesting categorial approach to
models which could be dual to non-Abelian gauge
theory~\cite{GrSc98}. It uses the same colouring of links and
plaquettes of the lattice with representations and intertwiners,
respectively, as our dual model does on the dual lattice in $d=3$ and
seems to enjoy many close similarities. However, the gauge degrees of
freedom proposed in~\cite{GrSc98} are not in a straight forward way a
symmetry of the dually transformed model as derived in this
paper. Nevertheless, the ideas developed there might prove useful for
a further study of the dual model.

We thank M.B.~Halpern for bringing the paper~\cite{Ba82b} to our
attention. In~\cite{Ba82b} a plaquette formulation of lattice gauge
theory is derived emphasizing the non-Abelian Bianchi identity. Using
the mathematical methods that we describe in
Sections~\ref{sect_prelimhopf} to~\ref{sect_prelimhaar}, it can be
shown that the approach of~\cite{Ba82b} can be extended to a full
duality transformation similar to the treatment in our
Section~\ref{sect_transform}. The result would be the same as given in
Theorem~\ref{thm_dualpartition1}. Finally, we thank a referee for
drawing our attention to the paper~\cite{HoFa96} where a duality
transformation of classical Yang-Mills theory in continuous space-time
based on the loop approach is suggested.

This paper is organized as follows. In Section~\ref{sect_prelim}, we
recall all definitions and structures which are needed to present the
duality transformation. In particular these are the structure of
the Hopf algebra of representation functions $\Calg(G)$ of $G$ and the
lattice formulation of gauge theories.

In Section~\ref{sect_transform}, we present the duality transformation
in detail first for the partition function, and then for the
non-Abelian generalizations of the Wilson loop. We specialize our
result to the known cases of non-Abelian lattice gauge theory in $2$
dimensions and to Abelian lattice gauge theory in arbitrary dimension
in Section~\ref{sect_special}.

Finally, in Section~\ref{sect_outlook} we indicate how the dual
formulation can be used in conjunction with strong coupling expansion
techniques and discuss open questions and directions for further
research.

% ==============================================================================
%
\section{Preliminaries}
%
% ==============================================================================
\label{sect_prelim}

%-------------------------------------------------------------------------------
\subsection{The Hopf algebra of representation functions}
%-------------------------------------------------------------------------------
\label{sect_prelimhopf}

In this section we collect definitions and basic statements related to
the algebra of representation functions $\Calg(G)$ of $G$. These and
the results presented in the next section about the Peter-Weyl
decomposition and the Peter-Weyl theorem are basically text book
knowledge, see \eg~\cite{BrDi85,CaSe95}. We recall the basic facts to
fix our notation and present a purely algebraic treatment of the
relevant results which we have not found elsewhere in this form.

Let $G$ be a finite group or a compact Lie group. We denote
finite-dimensional complex vector spaces on which $G$ is represented
by $V_\rho$ and by $\rho\colon G\to\Aut V_\rho$ the corresponding
group homomorphisms. Since each finite-dimensional complex
representation of $G$ is equivalent to a unitary representation, we
select a set $\Rep$ containing one unitary representation of $G$ for
each equivalence class of finite-dimensional representations. The
tensor product, the direct sum and taking the dual are supposed to be
closed operations on this set. This amounts to a particular choice of
representation isomorphisms $\rho_1\otimes\rho_2\leftrightarrow\rho_3$
\etc, $\rho_j\in\Rep$, which is implicit in our formulas. We
furthermore denote by $\Irrep\subseteq\Rep$ the subset of irreducible
representations.

If $\rho\in\Rep$, we write $\rho^\ast$ for the dual representation and
denote the dual vector space of $V_\rho$ by $V_\rho^\ast$. The dual
representation is given by $\rho^\ast\colon G\mapsto \Aut
V_\rho^\ast$, where
\begin{equation}
\label{eq_dualrep}
  \rho^\ast(g)\colon V_\rho^\ast\to V_\rho^\ast,\quad
    \eta\mapsto \eta\circ\rho(g^{-1}),
\end{equation}
\ie\ $(\rho^\ast(g)\eta)(v)=\eta(\rho(g^{-1})v)$ for all $v\in
V_\rho$.

For the unitary representations $V_\rho$, $\rho\in\Rep$, we have
standard (sesquilinear) scalar products $\left<\cdot;\cdot\right>$ and
orthonormal bases $(v_j)$ in such a way that the basis $(v_j)$ of
$V_\rho$ is dual to the basis $(\eta^j)$ of $V_\rho^\ast$, \ie\
$\eta^j(v_k)=\delta_{jk}$. This means that duality is given by the
scalar product,
\begin{equation}
  \left<v_j;v_k\right>=\eta^j(v_k),\qquad
  \bigl<\eta^j;\eta^k\bigr>=\eta^k(v_j),
  \qquad 1\leq j,k\leq\dim V_\rho.
\end{equation}
There exists a one-dimensional 'trivial' representation of $G$ which
we denote by $V_{[1]}\cong\C$.

The functions 
\begin{equation}
  t_{\eta,v}^{(\rho)}\colon G\to\C,\quad g\mapsto\eta(\rho(g)v),
\end{equation}
where $\rho\in\Rep$, $v\in V_\rho$ and $\eta\in V_\rho^\ast$,
are called the \emph{representation functions} of $G$. They form a 
commutative and associative unital algebra
over $\C$,
\begin{equation}
  \Calg(G) := \{\,t_{\eta,v}^{(\rho)}\colon\quad
    \rho\in\Rep, v\in V_\rho, \eta\in V_\rho^\ast\,\},
\end{equation}
whose operations are given by
\begin{mathletters}
\begin{eqnarray}
\label{eq_operationsum}
  (t_{\eta,v}^{(\rho)} + t_{\theta,w}^{(\sigma)})(g)
    &:=& t_{\eta+\theta,v+w}^{(\rho\oplus\sigma)}(g),\\
\label{eq_operationprod}
  (t_{\eta,v}^{(\rho)}\cdot t_{\theta,w}^{(\sigma)})(g)
    &:=& t_{\eta\otimes\theta,v\otimes w}^{(\rho\otimes\sigma)}(g),
\end{eqnarray}%
\end{mathletters}%
where $\rho,\sigma\in\Rep$ and $v\in V_\rho$, $w\in V_\sigma$,
$\eta\in V_\rho^\ast$, $\theta\in V_\sigma^\ast$ and $g\in G$. The
zero element of $\Calg(G)$ is given by $t_{0,0}^{[1]}(g)=0$ and its
unit element by $t_{\eta,v}^{[1]}(g)=1$ where we have normalized
$\eta(v)=1$.

The algebra $\Calg(G)$ is furthermore equipped with a Hopf algebra
structure with the coproduct
$\Delta\colon\Calg(G)\to\Calg(G)\otimes\Calg(G)\cong\Calg(G\times G)$,
the co-unit $\epsilon\colon\Calg(G)\to\C$ and the antipode
$S\colon\Calg(G)\to\Calg(G)$ which are defined by
\begin{mathletters}
\begin{eqnarray}
\label{eq_matrixcopro}
  \Delta t_{\eta,v}^{(\rho)} (g,h) 
    &:=& t_{\eta,v}^{(\rho)}(g\cdot h),\\
  \epsilon t_{\eta,v}^{(\rho)}
    &:=& t_{\eta,v}^{(\rho)}(1),\\
\label{eq_matrixanti}
  S t_{\eta,v}^{(\rho)} (g) 
    &:=& t_{\eta,v}^{(\rho)}(g^{-1}),
\end{eqnarray}%
\end{mathletters}%
where $\rho\in\Rep$ and $v\in V_\rho$, $\eta\in V_\rho^\ast$ and
$g,h\in G$.

Since $G$ is a finite group or a compact Lie group, all
finite-dimensional representations of $G$ are completely
reducible. Moreover, all representations of $G\times G$ are tensor
products of representations of $G$ such that we have an isomorphism of
algebras $\Calg(G\times G)\cong\Calg(G)\otimes\Calg(G)$ which is used
in the definition of the coproduct. This tensor product is algebraic,
and there is no need for a topology or a completion of the tensor
product at this point.

In the standard orthonormal bases, the representation functions are
given by the coefficients of representation matrices,
\begin{equation}
  t_{mn}^{(\rho)}(g) := t_{\eta^m,v_n}^{(\rho)}(g) 
    = \eta^m(\rho(g)v_n) = \left<v_m;\rho(g)v_n\right> = {\rho(g)}_{mn},
\end{equation}
such that the coproduct corresponds to the matrix product,
\begin{equation}
%\label{eq_matrixcopro}
  \Delta t_{mn}^{(\rho)}(g,h) 
    = \sum_{j=1}^{\dim V_\rho} t_{mj}^{(\rho)}(g)t_{jn}^{(\rho)}(h),
\end{equation}
while the antipode refers to the inverse matrix, $S
t_{mn}^{(\rho)}(g)={({\rho(g)}^{-1})}_{mn}$, and the co-unit describes
the coefficients of the unit matrix, $\epsilon
t_{mn}^{(\rho)}=\delta_{mn}$.  Furthermore, the antipode relates a
representation with its dual,
\begin{eqnarray}
\label{eq_antipode}
  S t_{mn}^{(\rho)}(g) = \eta^m({\rho(g)}^{-1}v_n) 
    = (\rho^\ast(g)\eta^m)(v_n)
    = \left<\eta^n;\rho^\ast(g)\eta^m\right> 
    = t_{nm}^{(\rho^\ast)}(g),
\end{eqnarray}
which is just the conjugate representation because on the other hand
\begin{equation}
  S t_{mn}^{(\rho)}(g) = \left<v_m;\rho(g^{-1})v_n\right>
    = \left<\rho(g)v_m;v_n\right> 
    = \overline{\left<v_n;\rho(g)v_m\right>} 
    = \overline{t_{nm}^{(\rho)}(g)}.
\end{equation}
Here the bar denotes complex conjugation.

%-------------------------------------------------------------------------------
\subsection{Peter-Weyl decomposition and theorem}
%-------------------------------------------------------------------------------

The structure of the algebra $\Calg(G)$ can be understood if
$\Calg(G)$ is considered as a representation of $G\times G$ by
combined left and right translation of the function argument,
\begin{equation}
  (G\times G)\times\Calg(G)\to\Calg(G),\quad
    (g_1,g_2,f)\mapsto (h\mapsto f(g_1^{-1}hg_2)).
\end{equation}
It can then be decomposed into its irreducible components as a
representation of $G\times G$:

\begin{theorem}[Peter-Weyl Decomposition]
Let $G$ be a finite group or a compact Lie group. 
\begin{myenumerate}
\item 
  There is an isomorphism
\begin{equation}
\label{eq_structure_calg}
  \Calg(G)\cong_{G\times G} 
    \bigoplus_{\rho\in\Irrep}(V_\rho^\ast\otimes V_\rho),
\end{equation}
  of representations of $G\times G$. Here the direct sum is over one
  unitary representative of each equivalence class of
  finite-dimensional irreducible representations of $G$. The direct
  summands $V_\rho^\ast\otimes V_\rho$ are irreducible as representations
  of $G\times G$. 
\item
  The direct sum in~\eqref{eq_structure_calg} is orthogonal with
  respect to the $L^2$ scalar product on $\Calg(G)$ which is formed
  using the Haar measure on $G$ on the left hand side, and using the
  standard scalar products on the right hand side, namely
\begin{equation}
\label{eq_l2measure}
  {\bigl<t_{\eta,v}^{(\rho)};t_{\theta,w}^{(\sigma)}\bigr>}_{L^2}
    = \int_G\overline{t_{\eta,v}^{(\rho)}(g)}\cdot t_{\theta,w}^{(\sigma)}(g)\,dg
    = \frac{1}{\dim V_{\rho}}\delta_{\rho\sigma}
        \overline{\left<\eta;\theta\right>}\left<v;w\right>,
\end{equation}
  where $\rho,\sigma\in\Irrep$ are irreducible. The Haar measure is
  denoted by $\int_G\,dg$ and normalized such that $\int_G\,dg=1$.
\end{myenumerate}
\end{theorem}

\begin{remark}
\begin{myenumerate}
\item
  If $G$ is finite, the Haar measure coincides with the normalized
  summation over all group elements.
\item
  The decomposition~\eqref{eq_structure_calg} directly corresponds to
  our notation of the representation functions $t^{(\rho)}_{nm}(g)$ if
  $\rho\in\Irrep$ is irreducible.
\item
  Each representation function $f\in\Calg(G)$ has a decomposition
  according to~\eqref{eq_structure_calg},
\begin{equation}
\label{eq_peterweyl_series}
  f = \sum_{\rho\in\Irrep}f_\rho,
\end{equation}
  such that we find for the $L^2$-norm
\begin{equation}
  {||f||}^2_{L^2} = \sum_{\rho\in\Irrep}\frac{1}{\dim V_\rho}{||f_\rho||}^2,
\end{equation}
  where $f_\rho\in V_\rho^\ast\otimes V_\rho$, $\rho\in\Irrep$, and
  all except finitely many $f_\rho$ are zero. Here ${||f_\rho||}^2$ is
  the trace norm for the finite-dimensional space
  $V_\rho^\ast\otimes V_\rho\cong\End V_\rho$.
\end{myenumerate}%
\end{remark}%

\noindent
The analytical aspects of $\Calg(G)$ can now be stated. 

\begin{theorem}[Peter-Weyl Theorem]
Let $G$ be a compact Lie group. Then $\Calg(G)$ is dense in $L^2(G)$.
\end{theorem}

\begin{remark}
\begin{myenumerate}
\item
  We use the Peter-Weyl theorem to complete $\Calg(G)$ with respect to
  the $L^2$ norm to $L^2(G)$. Functions $f\in L^2(G)$ then correspond
  to square summable series in~\eqref{eq_peterweyl_series}. These
  series are thus invariant under a reordering of summands, and their
  limits commute with group integrations. We will make use of these
  invariances in the duality transformation.
\item
  If $G$ is a finite group, $\Calg(G)$ is a finite-dimensional vector
  space such that the corresponding results hold trivially.
\end{myenumerate}
\end{remark}

%-------------------------------------------------------------------------------
\subsection{Character decomposition}
%-------------------------------------------------------------------------------

If $G$ is a finite group or a compact Lie group, the characters
$\chi^{(\rho)}\colon G\to\C$ associated with the finite-dimensional
unitary representations $\rho\in\Rep$ of $G$ are obtained from the
representation functions by
\begin{equation}
  \chi^{(\rho)} := \sum_{j=1}^{\dim V_\rho} t_{jj}^{(\rho)}.
\end{equation}
Each class function $f\in\Calg(g)$ has a character decomposition
\begin{equation}
\label{eq_chardecomp}
  f(g) = \sum_{\rho\in\Irrep}\chi^{(\rho)}(g)\,\hat f_\rho,
    \qquad\mbox{where}\qquad 
  \hat f_\rho = \dim V_\rho\,\int_G\overline{\chi^{(\rho)}(g)}f(g)\,dg.
\end{equation}
The completion of $\Calg(G)$ to $L^2(G)$ is compatible with this
decomposition.

%-------------------------------------------------------------------------------
\subsection{Projector description of the Haar measure}
%-------------------------------------------------------------------------------
\label{sect_prelimhaar}

For the duality transformation, it is important to understand the Haar
measure on $G$ in the picture of the Peter-Weyl
decomposition~\eqref{eq_structure_calg}. We describe the Haar measure
in terms of projectors.

\begin{proposition}
Let $G$ be a finite group or a compact Lie group and $\rho\in\Rep$ be
a finite-dimensional unitary representation of $G$ with the
orthogonal decomposition
\begin{equation}
  V_\rho\cong\bigoplus_{j=1}^k V_{\tau_j},\qquad \tau_j\in\Irrep, k\in\N,
\end{equation}
into irreducible components $\tau_j$. Let $P^{(j)}\colon V_\rho\to
V_{\tau_j}\subseteq V_\rho$ be the $G$-invariant orthogonal projectors
associated with the above decomposition. Assume that precisely the
first $\ell$ components $\tau_1,\ldots,\tau_\ell$, $0\leq\ell\leq k$,
are equivalent with the trivial representation. Then the Haar measure
of a representation function $t_{mn}^{(\rho)}$, $1\leq m,n\leq\dim
V_\rho$, is given by
\begin{equation}
\label{eq_haaralg}
  \int_G t_{mn}^{(\rho)}(g)\,dg = \sum_{j=1}^\ell\bigl<v_m;P^{(j)}v_n\bigr>
    = \sum_{j=1}^\ell\bigl<P^{(j)}v_m;P^{(j)}v_n\bigr> 
    = \sum_{j=1}^\ell P^{(j)}_mP^{(j)}_n,
\end{equation}
where $P^{(j)}_m=\eta(P^{(j)}v_m)$ denotes the matrix elements of the
$j$-th projector. Here $\eta\in V_\rho^\ast$ is the normalized linear
form which is zero everywhere except on the one-dimensional
sub-spaces $V_{\tau_j}\subseteq V_\rho$, $1\leq j\leq\ell$. 
\end{proposition}

\begin{proof}
The representation function is Peter-Weyl decomposed by inserting
$\openone=\sum_{j=1}^k P^{(j)}$ twice into the right hand side of
$t_{mn}^{(\rho)}(g)=\left<v_m;\rho(g)v_n\right>$. We use hermiticity
${P^{(j)}}^\dagger=P^{(j)}$, $G$-invariance $[P^{(j)},\rho(g)]=0$ and
transversality $P^{(i)}P^{(j)}=\delta_{ij}P^{(j)}$ to obtain
\begin{equation}
  \int_G t_{mn}^{(\rho)}(g)\,dg =
  \sum_{j=1}^k\int_G\bigl<P^{(j)}v_m;\rho(g)P^{(j)}v_n\bigr>\,dg.
\end{equation}
Since the Haar measure is bi-invariant, all terms vanish except
those corresponding to the $\tau_j$, $1\leq j\leq\ell$, which are
equivalent to the trivial representation.
\end{proof}

%-------------------------------------------------------------------------------
\subsection{The Lattice formulation of non-Abelian gauge theories}
%-------------------------------------------------------------------------------

The purpose of this section is to fix a notation in which we can write
down the partition function and expectation values of non-Abelian
lattice gauge theory and which is suitable to formulate the duality
transformation. For all other issues we refer the reader to the
standard text books on lattice gauge theory, \eg~\cite{Ro92,MoMu94},
and references therein.

We consider a regular hyper-cubic lattice corresponding to an
Euclidean space-time of dimension $d\geq 2$. The \emph{lattice points}
(vertices) are denoted by tuples of integer numbers
\begin{equation}
 \Lambda^0 := \{\,(i_1,\ldots,i_d)\in\Z^d\colon\quad 
                i_\mu\in\{1,\ldots,N_\mu\}\,\},
\end{equation}
where the lattice is of size $N_\mu$ in the $\mu$-th dimension. We denote
the unit vectors along the lattice axes by
\begin{equation}
  \hat\mu:=(0,\ldots,0,\underbrace{1}_\mu,0,\ldots,0),\qquad 1\leq\mu\leq d.
\end{equation}
Thus $i+\hat\mu$ refers to the neighbour of the point $i$ in the
direction $\mu$. We choose periodic (\ie\ toroidal) boundary
conditions and identify $i\pm N_\mu\cdot\hat\mu\equiv i$ for all
$\mu\in\{1,\ldots,d\}$. It is crucial for the existence of various
integrals that we work on a finite lattice. Periodic boundary
conditions are the most convenient choice for our purpose. The
non-trivial homologies introduced by the periodic boundary conditions
do not play any role for the duality transformation in the form
presented below.

The set of all \emph{links} (edges) is called $\Lambda^1$, the
set of all \emph{plaquettes} (faces, squares) $\Lambda^2$, and
more generally the set of all $k$-\emph{cells} $\Lambda^k$. These are
specified by
\begin{equation}
 \Lambda^k := \{\,(i,\mu_1,\ldots,\mu_k)\colon\quad i\in\Lambda^0,
                   1\leq\mu_1<\cdots<\mu_k\leq d\,\}.
\end{equation}
In particular, the sets $\Lambda^k$, $0\leq k\leq d$, are all finite.
The $k$-cells are considered unoriented, \eg\ we do not want to
distinguish the plaquette $(i,\mu,\nu)$ from $(i,\nu,\mu)$. In our
notation both are represented in the standard way $(i,\mu,\nu)$ where
$\mu<\nu$.

The \emph{configurations} of lattice gauge theory are the maps
\begin{equation}
  g\colon\Lambda^1\to G,\quad (i,\mu)\mapsto g_{i\mu},
\end{equation}
which assign a group element $g_{i\mu}\in G$ to each link
$(i,\mu)\in\Lambda^1$ of the lattice. These group elements correspond
to the parallel transports of the gauge connection along the links.

\begin{figure}[t]
\begin{center}
\input{fig/plaquette_2.pictex}
\mycaption{fig_plaquette}{%
  The oriented product of link variables around a given plaquette
  $(i,\mu,\nu)\in\Lambda^2$ as given by~\eqref{eq_plaquette}.}
\end{center}
\end{figure}

The path integral measure depends on these configurations only via the
\emph{plaquette product} (see also Figure~\ref{fig_plaquette}),
\begin{equation}
\label{eq_plaquette}
  dg\colon\Lambda^2\to G,\quad
    (i,\mu,\nu)\mapsto dg_{i\mu\nu}:=
    g_{i\mu}\cdot g_{i+\hat\mu,\nu}\cdot g_{i+\hat\nu,\mu}^{-1}\cdot g_{i,\nu}^{-1},
\end{equation}
which is the path ordered product of the link variables around a given
plaquette $(i,\mu,\nu)\in\Lambda^2$. 

For arbitrary \emph{generating functions} $\phi\colon\Lambda^0\to G$,
any class function of $G$, evaluated on $dg_{i\mu\nu}$, is invariant
under the \emph{gauge transformation}
\begin{equation}
\label{eq_regauge}
  g_{i\mu}\mapsto g_{i\mu}^\prime:=\phi_i\cdot g_{i\mu}\cdot\phi_{i+\hat\mu}^{-1}.
\end{equation}
Let $G$ be a compact Lie group (or a finite group) and $\int_G\,dg$
denote the Haar measure. The path integral integrates over all
configurations, \ie\ it consists of one integration over $G$ for each
link. We denote this integration by
\begin{equation}
  \int\sym{D}g = \Bigl(\prod_{(i,\mu)\in\Lambda^1}\int_G\,dg_{i\mu}\Bigr)
    := \underbrace{\int_G\,dg_{i\mu}\cdots\int_G\,dg_{i\mu}}_{
       \mbox{one $\int_G$ for each link}}.
\end{equation}
The path integral measure of lattice gauge theory is this integration
together with a Boltzmann weight $\exp(-S(dg))$. Here the (local)
\emph{action} $S$ is given by a sum over all plaquettes,
\begin{equation}
  S(dg) := \sum_{(i,\mu,\nu)\in\Lambda^2} s(dg_{i\mu\nu}),
\end{equation} 
where $s\colon G\to\R$ is an $L^2$ integrable class function on $G$
which is bounded from below. The action is thus manifestly gauge
invariant.

The full \emph{partition function} of lattice gauge theory with gauge
group $G$ finally reads
\begin{equation}
\label{eq_partition}
  Z = \int\sym{D}g\,\exp(-S(dg))
    = \Bigl(\prod_{(i,\mu)\in\Lambda^1}\int_G\,dg_{i\mu}\Bigr)\,
      \prod_{(i,\mu,\nu)\in\Lambda^2}f(dg_{i\mu\nu}),
\end{equation}
where $f(g)=\exp(-s(g))$ is a positive real and $L^2$ integrable
class function on $G$. The standard example for the action is the
\emph{Wilson action},
\begin{equation}
  s(g) = -\frac{\beta}{2\dim V_\rho}(\chi^{(\rho)}(g)+\overline{\chi^{(\rho)}(g)}),
\end{equation}
where $\chi^{(\rho)}\colon G\to\C$ is the character of a unitary
matrix representation $\rho$ of $G$, usually the fundamental
representation. The \emph{inverse temperature} $\beta$ encodes the
coupling constant.

In~\eqref{eq_partition} and in the following we are careful not to
waste letters of the alphabet for dummy indices. The indices $i,\mu$
of the first product sign are just there to indicate that group
integration is performed for each link of the lattice. We adopt the
convention that $i$ and $\mu$ in this case do not have any meaning
outside the enclosing brackets. So we can use the same letters again
after the closing bracket.

%-------------------------------------------------------------------------------
\subsection{Gauge invariant quantities}
%-------------------------------------------------------------------------------

Of course, Wilson loops are gauge invariant expressions. However, in
non-Abelian lattice gauge theory, not all gauge invariant expressions
are given by Wilson loops. The generic gauge invariant expressions are
so-called \emph{spin networks} which include branchings of the Wilson
lines. This is familiar, for example, from the expression which is
used to determine the static three-quark potential.

Here we formalize this generalization and give the following slightly
technical definition:
\begin{definition}
\label{def_spinnetwork}
Let $\tau\colon\Lambda^1\to\Irrep,(j,\kappa)\mapsto\tau_{j\kappa}$,
associate a finite-dimensional irreducible unitary representation of
$G$ with each link of the lattice. Choose furthermore for each lattice
point $j\in\Lambda^0$ an intertwiner
\begin{equation}
\label{eq_spinnetmap}
  Q^{(j)}\colon\bigotimes_{\mu=1}^d\tau_{j-\hat\mu,\mu}\to
    \bigotimes_{\mu=1}^d\tau_{j\mu},
\end{equation}
which maps from the tensor product of the representations of the $d$
`incoming' links to the tensor product of the $d$ `outgoing' links at
the point $j\in\Lambda^0$. 

The \emph{non-Abelian Wilson loop} (or \emph{spin network}) associated with
$\tau$ and $Q^{(j)}$ is defined by
\begin{eqnarray}
\label{eq_spinnetwork}
  \Wloop 
  &:=& \Bigl(\prod_{(j,\kappa)\in\Lambda^1}\sum_{a_{j\kappa},b_{j\kappa}}\Bigr)\,
  \Bigl(\prod_{(j,\kappa)\in\Lambda^1}
        t_{a_{j\kappa}b_{j\kappa}}^{(\tau_{j\kappa})}(g_{j\kappa})\Bigr)\\
  &\times&\Bigl(\prod_{j\in\Lambda^0}
    Q^{(j)}_{(b_{j-\hat1,1}\ldots b_{j-\hat d,d}),(a_{j1}\ldots a_{jd})}\Bigr).\nn
\end{eqnarray}
The indices $(b_{j-\hat1,1}\ldots b_{j-\hat d,d})$ of $Q^{(j)}$ refer
to the tensor factors of the domain of $Q^{(j)}$ (`incoming')
while the $(a_{j1}\ldots a_{jd})$ refer to the image
(`outgoing'), \cf~\eqref{eq_spinnetmap}.
\end{definition}

In~\eqref{eq_spinnetwork} we have abbreviated the summation of the
vector indices $a_{j\kappa}$ and $b_{j\kappa}$ for all links by
\begin{equation}
  \Bigl(\prod_{(j,\kappa)\in\Lambda^1}\sum_{a_{j\kappa},b_{j\kappa}}\Bigr) 
  = \underbrace{\sum_{a_{j\kappa},b_{j\kappa}=1}^{\dim V_{\tau_{j\kappa}}}
          \cdots\sum_{a_{j\kappa},b_{j\kappa}=1}^{\dim V_{\tau_{j\kappa}}}}_{
       \mbox{one $\sum$ for each link}}.
\end{equation}
This notation is frequently used in the duality transformation in
Section~\ref{sect_transform}.

\begin{proposition}
  The expression $\Wloop$ of the non-Abelian Wilson loop
  in~\eqref{eq_spinnetwork} is gauge invariant under the
  transformation~\eqref{eq_regauge}.
\end{proposition}

\begin{proof}
  Consider an arbitrary lattice point $j\in\Lambda^0$. The gauge
  transformation~\eqref{eq_regauge} multiplies all `incoming' links by
  $\phi_j^{-1}$ and all `outgoing' links by $\phi_j$. Since $Q^{(j)}$
  in~\eqref{eq_spinnetmap} is an intertwiner, $\Wloop$ is unchanged. This
  holds for all lattice points $j\in\Lambda^0$.
\end{proof}

\begin{remark}
\begin{myenumerate}
\item Depending on the representations $\tau_{j\kappa}$, there are situations
  in which for some $j\in\Lambda^0$ the only choice is $Q^{(j)}=0$ and thus
  $\Wloop=0$. This is the case \eg\ if a would-be Wilson loop is not properly
  closed.
\item All links $(j,\kappa)$ for which $\tau_{j\kappa}\cong V_{[1]}$
  is the trivial representation, disappear from the
  expression~\eqref{eq_spinnetwork}.  For an ordinary Wilson loop, for
  example, all links are labelled with the trivial representation
  except those links which are part of the loop. These are labelled
  with the fundamental representation of $G$. The intertwiners
  $Q^{(j)}$ (if non-vanishing) are in this case uniquely determined up
  to normalization.
\item In Definition~\ref{def_spinnetwork}, the requirement that the
  $\tau_{j\kappa}$ be irreducible, and that there be only one factor
  $t_{ab}^{(\tau)}$ per link, can be imposed without loss of
  generality. Otherwise the expression $\Wloop$ would decompose
  into similar expressions involving only irreducible representations
  and only one function $t_{ab}^{(\tau)}$ per link,
  \cf~\eqref{eq_operationsum} and~\eqref{eq_operationprod}.
\item If $G$ is Abelian, then $\Irrep\cong\Z$, and all irreducible
  representations are one-dimensional. Thus $\Wloop$ can be decomposed
  into a sum of products of Abelian Wilson loops.
\end{myenumerate}
\end{remark}

\noindent
The normalized expectation value of a non-Abelian Wilson loop finally
reads 
\begin{equation}
\label{eq_expect}
  \bigl<\Wloop\bigr>=\frac{1}{Z}\int\sym{D}g\,\Wloop\,\exp(-S(dg)),
\end{equation}
\cf~\eqref{eq_partition} and~\eqref{eq_spinnetwork}.

% ==============================================================================
%
\section{The duality transformation}
%
% ==============================================================================
\label{sect_transform}

In this section we present the duality transformation in detail. We
start with the transformation of the partition function and then turn
to the expectation value of the non-Abelian Wilson loop.

%-------------------------------------------------------------------------------
\subsection{The dual of the partition function}
%-------------------------------------------------------------------------------
\label{sect_transformpart}

We start with the partition function of non-Abelian lattice gauge
theory~\eqref{eq_partition}. Since the Boltzmann weight function $f(g)$ is an
$L^2$ class function, we can insert its character decomposition,
see~\eqref{eq_chardecomp}:
\begin{equation}
\label{eq_charexp}
  f(g) = \sum_{\rho\in\Irrep}\hat f_\rho\chi^{(\rho)}(g)
  = \sum_{\rho\in\Irrep}\hat f_\rho\sum_{n=1}^{\dim V_\rho}t_{nn}^{(\rho)}(g).
\end{equation}
The partition function thus reads
\begin{equation}
  Z = \Bigl(\prod_{(i,\mu)\in\Lambda^1}\int_G\,dg_{i\mu}\Bigr)\,
      \prod_{(i,\mu,\nu)\in\Lambda^2}\Bigl(\sum_{\rho_{i\mu\nu}\in\Irrep}
      \hat f_{\rho_{i\mu\nu}}\sum_{n_{i\mu\nu}=1}^{\dim V_{\rho_{i\mu\nu}}}
        t_{n_{i\mu\nu}n_{i\mu\nu}}^{(\rho_{i\mu\nu})}
          (g_{i\mu}g_{i+\hat\mu,\nu}g_{i+\hat\nu,\mu}^{-1}g_{i,\nu}^{-1})\Bigr),
\end{equation}
where each plaquette $(i,\mu,\nu)\in\Lambda^2$ is coloured with an irreducible
representation $\rho_{i\mu\nu}\in\Irrep$, and the indices $n_{i\mu\nu}$ which
originate from the traces, are summed once for each plaquette.

The next step is to employ the coproduct and the antipode,
see~\eqref{eq_matrixcopro} and~\eqref{eq_matrixanti}, in order to remove all
group products and inverses from the arguments of the representation
functions.  Furthermore, we reorganize the summations: 
\begin{eqnarray}
\label{eq_step2}
  Z&=&\Bigl(\prod_{(i,\mu)\in\Lambda^1}\int_G\,dg_{i\mu}\Bigr)\,
    \Bigl(\prod_{(i,\mu,\nu)\in\Lambda^2}\sum_{\rho_{i\mu\nu}\in\Irrep}\Bigr)\,
    \prod_{(i,\mu,\nu)\in\Lambda^2}\Biggl[\hat f_{\rho_{i\mu\nu}}\,\times\Biggr.\\
  &\times&\Biggl.\Bigl(\sum_{n_{i\mu\nu},m_{i\mu\nu},p_{i\mu\nu},q_{i\mu\nu}}
    t_{n_{i\mu\nu}m_{i\mu\nu}}^{(\rho_{i\mu\nu})}(g_{i\mu})
    t_{m_{i\mu\nu}p_{i\mu\nu}}^{(\rho_{i\mu\nu})}(g_{i+\hat\mu,\nu})
    St_{p_{i\mu\nu}q_{i\mu\nu}}^{(\rho_{i\mu\nu})}(g_{i+\hat\nu,\mu})
    St_{q_{i\mu\nu}n_{i\mu\nu}}^{(\rho_{i\mu\nu})}(g_{i\nu})\Bigr)\Biggr].\nn
\end{eqnarray}
In all places where we have applied the coproduct, written
schematically as
\begin{equation*}
  t_{nn}(g_1g_2g_3g_4)=\sum_{m,p,q}t_{nm}(g_1)t_{mp}(g_2)t_{pq}(g_3)t_{qn}(g_4),
\end{equation*}
new vector indices have entered which are associated with the
plaquette $(i,\mu,\nu)\in\Lambda^2$ and denoted by $m_{i\mu\nu}$,
$p_{i\mu\nu}$ and $q_{i\mu\nu}$. They are summed over the range
$1\ldots\dim V_{\rho_{i\mu\nu}}$. 

\begin{figure}[t]
\begin{center}
\input{fig/indices_2.pictex}
\mycaption{fig_linkplaqu}{% 
  All plaquettes whose boundary contains a given link
  $(i,\mu)\in\Lambda^1$. This figure shows the situation in $d=3$. In
  generic dimension $d$, there are $2(d-1)$ such plaquettes. The
  coordinate axes are chosen such that $1\leq\lambda<\mu<\nu\leq d$. 
  We illustrate the numbering of the lattice points $i$,
  $i-\hat\lambda$, \etc, which are needed to describe the relevant
  plaquettes $(i-\hat\lambda,\lambda,\mu)$, $(i,\mu,\nu)$, \etc\ with
  their correct orientations. Furthermore the letters $n$, $m$, $p$,
  $q$ indicate to which lattice point the vector index summations
  $n_{i\mu\nu}$, $m_{i\mu\nu}$, $p_{i\mu\nu}$ and $q_{i\mu\nu}$ of a
  given plaquette $(i,\mu,\nu)$ belong.}
\end{center}
\end{figure}

In order to perform the group integrations, we have to reorganize the
product in~\eqref{eq_step2} such that all representation functions
whose argument refers to the same link are grouped together.
Therefore we have to find all plaquettes which contain a given link
$(i,\mu)\in\Lambda^1$ in their boundary, \ie\ all plaquettes which
\emph{cobound} the link. Figure~\ref{fig_linkplaqu} illustrates this
situation for $d=3$. The reorganized product reads
\begin{eqnarray}
\label{eq_reordering}
  Z&=&\Bigl(\prod_{(i,\mu,\nu)\in\Lambda^2}\sum_{\rho_{i\mu\nu}\in\Irrep}\Bigr)\,
    \Bigl(\prod_{(i,\mu,\nu)\in\Lambda^2}\hat
    f_{\rho_{i\mu\nu}}\Bigr)\,
  \Bigl(\prod_{(i,\mu,\nu)\in\Lambda^2}
    \sum_{n_{i\mu\nu},m_{i\mu\nu},p_{i\mu\nu},q_{i\mu\nu}}\Bigr)\\
  &\times&\prod_{(i,\mu)\in\Lambda^1}\Biggl\{\int_G\,dg_{i\mu}\,\Biggl[
    \prod_{\lambda=1}^{\mu-1}\biggl(
      t_{m_{i-\hat\lambda,\lambda,\mu}p_{i-\hat\lambda,\lambda,\mu}}^{
         (\rho_{i-\hat\lambda,\lambda,\mu})}(g_{i\mu})\cdot
      St_{q_{i\lambda\mu}n_{i\lambda\mu}}^{(\rho_{i\lambda\mu})}(g_{i\mu})\biggr)
      \Biggr.\Biggr.\nn\\
  &&\qquad\Biggl.\Biggl.\times\prod_{\nu=\mu+1}^d\biggl(
      St_{p_{i-\hat\nu,\mu,\nu}q_{i-\hat\nu,\mu,\nu}}^{(\rho_{i-\hat\nu,\mu,\nu})}(g_{i\mu})
      \cdot t_{n_{i\mu\nu}m_{i\mu\nu}}^{(\rho_{i\mu\nu})}(g_{i\mu})\biggr)
    \Biggr]\Biggr\}.\nn
\end{eqnarray}

The expression~\eqref{eq_reordering} means that each plaquette is
coloured with an irreducible representation. There is furthermore a
(dual Boltzmann) weight factor $\hat f_\rho$ per plaquette. Since we
have reorganized the product of the representation functions
$t_{ij}(g)$ such that those whose argument refers to the same link
$(i,\mu)\in\Lambda^1$ are placed next to each other, the group
integrations in~\eqref{eq_reordering} are performed for each link
separately. In the integrand, the two products $\prod_{\lambda}$ and
$\prod_{\nu}$ enumerate all plaquettes cobounding the link $(i,\mu)$
in arbitrary dimension $d$ and are such that $1\leq\lambda<\mu<\nu\leq
d$ always. In dimension $d$, there are $2(d-1)$ factors for each link
direction $\mu$.

Next we eliminate the antipodes using~\eqref{eq_antipode}. The group
integrals in~\eqref{eq_reordering} thus read
\begin{eqnarray}
\label{eq_delantipode}
  \int_G\,dg_{i\mu}\Biggl[
    \prod_{\lambda=1}^{\mu-1}\biggl(
      t_{m_{i-\hat\lambda,\lambda,\mu}p_{i-\hat\lambda,\lambda,\mu}}^{
         (\rho_{i-\hat\lambda,\lambda,\mu})}(g_{i\mu})\cdot
      t_{n_{i\lambda\mu}q_{i\lambda\mu}}^{(\rho_{i\lambda\mu}^\ast)}(g_{i\mu})\biggr)
      \Biggr.\nn\\
  \qquad\Biggl.\times\prod_{\nu=\mu+1}^d\biggl(
      t_{q_{i-\hat\nu,\mu,\nu}p_{i-\hat\nu,\mu,\nu}}^{(\rho_{i-\hat\nu,\mu,\nu}^\ast)}(g_{i\mu})
      \cdot t_{n_{i\mu\nu}m_{i\mu\nu}}^{(\rho_{i\mu\nu})}(g_{i\mu})\biggr)
    \Biggr].
\end{eqnarray}
Now the group integrations can be performed using the
expression~\eqref{eq_haaralg} in terms of projectors. We obtain
\begin{eqnarray}
\label{eq_step4}
  &\displaystyle\int_G\,dg_{i\mu}\biggl[\cdots\biggr]&\nn\\
   = &\displaystyle\sum_{P\in\sym{P}_{i\mu}}
     P_{(\underbrace{m_{i-\hat\lambda,\lambda,\mu}n_{i\lambda\mu}\ldots}_{
       \lambda\in\{1,\ldots,\mu-1\}})
        (\underbrace{q_{i-\hat\nu,\mu,\nu}n_{i\mu\nu}\ldots}_{
       \nu\in\{\mu+1,\ldots,d\}})}\cdot
     P_{(\underbrace{p_{i-\hat\lambda,\lambda,\mu}q_{i\lambda\mu}\ldots}_{
       \lambda\in\{1,\ldots,\mu-1\}})
        (\underbrace{p_{i-\hat\nu,\mu,\nu}m_{i\mu\nu}\ldots}_{
       \nu\in\{\mu+1,\ldots,d\}})}.&
\end{eqnarray}
Here the sum is over a complete set $\sym{P}_{i\mu}$ of inequivalent
orthogonal projectors onto the trivial one-dimensional components in
the decomposition of
\begin{equation}
\label{eq_step5}
  \underbrace{
    (\rho_{i-\hat\lambda,\lambda,\mu}\otimes\rho_{i\lambda\mu}^\ast)\otimes\cdots}_{
    \lambda\in\{1,\ldots,\mu-1\}}\otimes
  \underbrace{
    (\rho_{i-\hat\nu,\mu,\nu}^\ast\otimes\rho_{i\mu\nu})\otimes\cdots}_{
    \nu\in\{\mu+1,\ldots,d\}}
\end{equation}
into irreducible components. The dots ``$\cdots$'' indicate that there
are pairs $\rho\otimes\rho^\ast$ of tensor factors for all
$\lambda\in\{1,\ldots,\mu-1\}$ and pairs $\rho^\ast\otimes\rho$ for
all $\nu\in\{\mu+1,\ldots,d\}$. This gives the correct result in
arbitrary dimension and takes into account the orientation of the
link $(i,\mu)$ in the boundary of the given plaquette. Opposite
orientations of the link correspond to dual representations.
Similarly, the dots ``$\ldots$'' in~\eqref{eq_step4} indicate that
there is one pair of indices for each pair $\rho\otimes\rho^\ast$
resp.\ $\rho^\ast\otimes\rho$ which appears in~\eqref{eq_step5}.

With this step we have evaluated all group integrations over the links. As new
degrees of freedom for the dual path integral, the colourings of all
plaquettes with irreducible representations of $G$ have emerged:
\begin{eqnarray}
\label{eq_step6}
  Z&=&\Bigl(\underbrace{\prod_{(i,\mu,\nu)\in\Lambda^2}\sum_{\rho_{i\mu\nu}\in\Irrep}}_{
      \ontop{\mbox{part of the}}{\mbox{dual path integral}}}\Bigr)\,
    \Bigl(\underbrace{\prod_{(i,\mu,\nu)\in\Lambda^2}\hat
    f_{\rho_{i\mu\nu}}}_{\mbox{dual Boltzmann weight}}\Bigr)\,
  \Bigl(\underbrace{\prod_{(i,\mu,\nu)\in\Lambda^2}
    \sum_{n_{i,\mu,\nu},m_{i\mu\nu},p_{i\mu\nu},q_{i\mu\nu}}}_{
      \mbox{vector index summations}}\Bigr)\nn\\
  &\times&\prod_{(i,\mu)\in\Lambda^1}\biggl(
     \underbrace{\sum_{P\in\sym{P}_{i\mu}} P_{(\cdots)(\cdots)}\cdot
      P_{(\cdots)(\cdots)}}_{\mbox{one constraint for each link}}\biggr).
\end{eqnarray}
The last sum has the form as in~\eqref{eq_step4} for each link
$(i,\mu)$ appearing in the product.

There are still the summations over the vector indices
$n_{i\mu\nu},\ldots,q_{i\mu\nu}$ for all plaquettes. We call the
factors arising from them and from the projectors \emph{gauge
constraints} because they generalize the conditions which ensure in
the Abelian case that the integer $2$-form appearing in the dual
model, is co-closed. They seem to form a number of complicated
non-local constraints.  However, the summations can again be reordered
in a suitable way so that the constraints appear only locally in a
certain sense. In order to see this, some geometric intuition is
necessary.

The projectors which have appeared in the group integration
$\int_G\,dg_{i\mu}$ for the link $(i,\mu)\in\Lambda^1$ and their
indices are of the form
\begin{equation}
\label{eq_projsplit}
  P_{(m_{i-\hat\lambda,\lambda,\mu}n_{i\lambda\mu}\ldots)
     (q_{i-\hat\nu,\mu,\nu}n_{i\mu\nu}\ldots)}\cdot
  P_{(p_{i-\hat\lambda,\lambda,\mu}q_{i\lambda\mu}\ldots)
     (p_{i-\hat\nu,\mu,\nu}m_{i\mu\nu}\ldots)},
\end{equation}
\ie\ the indices correspond to the plaquettes located at
$i-\hat\lambda$, $i$, $\ldots$, $i-\hat\nu$, $i$, $\ldots$ for the
first projector and similarly for the second projector. 

\begin{figure}[t]
\begin{center}
\input{fig/indices2_2.pictex}
%epsfig{file=fig/indices2,scale=0.6} 
\mycaption{fig_indices2}{%
  All plaquettes containing a given lattice point $i\in\Lambda^0$ in
  their boundaries. This figure shows the situation in $d=3$. In the
  generic case, there are $2d(d-1)$ such plaquettes. This figure
  illustrates the numbering of points $i-\hat\mu-\hat\nu$,
  $i-\hat\nu$, $i-\hat\mu$ and $i$ which are used to specify the
  relevant plaquettes $(i-\hat\mu-\hat\nu,\mu,\nu)$,
  $(i-\hat\nu,\mu,\nu)$, $(i-\hat\mu,\mu,\nu)$ and $(i,\mu,\nu)$ in the 
  $(\mu,\nu)$-plane and similarly for the other directions,
  \cf~\eqref{eq_constraint}. The letters $n,m,p,q$ indicate which
  lattice points the vector index summations are associated with. The
  labelling for a given plaquette $(i,\mu,\nu)$ starts with $n$ at $i$
  and proceeds counter-clockwise with $m,p,q$.}
\end{center}
\end{figure}

The crucial geometrical observation (Figure~\ref{fig_linkplaqu}) is
that all vector indices
$m_{i-\hat\lambda,\lambda,\mu},n_{i\lambda\mu},\ldots$,
$q_{i-\hat\nu,\mu,\nu},n_{i\mu\nu},\ldots$ which appear at the first
projector correspond to the lattice point $i$ whereas all vector
indices $p_{i-\hat\lambda,\lambda,\mu},q_{i\lambda\mu},\ldots$,
$p_{i-\hat\nu,\mu,\nu},m_{i\mu\nu},\ldots$ of the second projector
correspond to the lattice point $i+\hat\mu$. Recall that the
enumeration of the indices $n,m,p,q$ for a given plaquette
$(i,\mu,\nu)\in\Lambda^2$ was done starting with $n$ at the point $i$,
then proceeding counter-clockwise in the $(\mu,\nu)$-plane. It is
thus possible to associate the summations over the $n,m,p,q$ with the
lattice points $i\in\Lambda^0$. For each link $(i,\mu)\in\Lambda^1$,
one of the two projectors in~\eqref{eq_step4} then belongs to $i$, the
other to $i+\hat\mu$. However, the projectors can be separated only if
the summation over the projectors $\sym{P}_{i\mu}$ which is associated
with a link rather than a point, can be removed from these
expressions.

In the partition function, there is in total one such summation over
projectors for each link of the lattice. It is thus natural to
consider these summations as part of the dual path integral. If this
is done, we can reorganize the expressions such that all vector index
summations and projectors are associated with the lattice points
$i\in\Lambda^0$ and such that they involve only data from the
neighbouring links and plaquettes. This is what is meant by
'locality'.  In Figure~\ref{fig_indices2} we show the lattice point
$i\in\Lambda^0$ together with the $2d$ links and $2d(d-1)$ plaquettes
which contain $i$. The partition function finally reads
\begin{equation}
\label{eq_step7}
  Z=\underbrace{\Bigl(\prod_{(i,\mu,\nu)\in\Lambda^2}\sum_{\rho_{i\mu\nu}\in\Irrep}\Bigr)\,
    \Bigl(\prod_{(i,\mu)\in\Lambda^1}\sum_{P^{(i\mu)}\in\sym{P}_{i\mu}}\Bigr)}_{
      \mbox{dual path integral}}\,
    \Bigl(\underbrace{\prod_{(i,\mu,\nu)\in\Lambda^2}\hat
    f_{\rho_{i\mu\nu}}}_{\mbox{dual Boltzmann weight}}\Bigr)\,
    \prod_{i\in\Lambda^0}\,C(i).
\end{equation}
where $C(i)$ encompasses the vector index summations and projectors associated
with the lattice point $i\in\Lambda^0$:
\begin{eqnarray}
\label{eq_constraint}
  C(i) &=& \Bigl(\prod_{1\leq\mu<\nu\leq d}
           \sum_{p_{i-\hat\mu-\hat\nu,\mu,\nu}=1}^{\dim V_{\rho_{i-\hat\mu-\hat\nu,\mu,\nu}}}
           \sum_{q_{i-\hat\nu,\mu,\nu}=1}^{\dim V_{\rho_{i-\hat\nu,\mu,\nu}}}
           \sum_{m_{i-\hat\mu,\mu,\nu}=1}^{\dim V_{\rho_{i-\hat\mu,\mu,\nu}}}
           \sum_{n_{i\mu\nu}=1}^{\dim V_{\rho_{i\mu\nu}}}\Bigr)\\
  &&\prod_{\mu=1}^d
     P^{(i\mu)}_{(\underbrace{m_{i-\hat\lambda,\lambda,\mu}n_{i\lambda\mu}\ldots}_{
       \lambda\in\{1,\ldots,\mu-1\}})
        (\underbrace{q_{i-\hat\nu,\mu,\nu}n_{i\mu\nu}\ldots}_{
       \nu\in\{\mu+1,\ldots,d\}})}\nn\\
  &&\qquad\cdot P^{(i-\hat\mu,\mu)}_{(\underbrace{p_{i-\hat\mu-\hat\lambda,\lambda,\mu}
                                                  q_{i-\hat\mu,\lambda,\mu}\ldots}_{
       \lambda\in\{1,\ldots,\mu-1\}})
        (\underbrace{p_{i-\hat\mu-\hat\nu,\mu,\nu}m_{i-\hat\mu,\mu,\nu}\ldots}_{
       \nu\in\{\mu+1,\ldots,d\}})}.\nn
\end{eqnarray}
The first product parameterizes all possible planes
by $\mu<\nu$. The four sums are then associated with the four
plaquettes $(i-\hat\mu-\hat\nu,\mu,\nu)$, $(i-\hat\nu,\mu,\nu)$,
$(i-\hat\mu,\mu,\nu)$ and $(i,\mu,\nu)$ in the $(\mu,\nu)$-plane which
contain the point $i$ (Figure~\ref{fig_indices2}). The last
product enumerates the `outgoing' and `incoming' links and contains
the projectors associated with this link and with the point $i$. Note
that the projectors in~\eqref{eq_constraint} for a given lattice point
$i\in\Lambda^0$ involve only vector indices whose summation is part of
the same $C(i)$.

The partition function~\eqref{eq_step7} consists now of a sum over
irreducible representations for all plaquettes and a sum over the
projectors onto the trivial components in the tensor
product~\eqref{eq_step5} for all links. This tensor product involves
the representations of all plaquettes which cobound the given link.
In particular, if the tensor product does not contain a trivial
component, this sum over projectors is empty and does not contribute
to the path integral. 

These results are summarized in the following theorem:

\begin{theorem}[Dual partition function]
\label{thm_dualpartition1}
Let $G$ be a compact Lie group or a finite group. The partition
function~\eqref{eq_partition} of lattice gauge theory with the gauge
group $G$ on a $d$-dimension\-al finite lattice with periodic boundary
conditions is equal to the expression
\begin{equation}
\label{eq_dualpartition1}
  Z=\underbrace{\Bigl(\prod_{(i,\mu,\nu)\in\Lambda^2}\sum_{\rho_{i\mu\nu}\in\Irrep}\Bigr)\,
    \Bigl(\prod_{(i,\mu)\in\Lambda^1}\sum_{P^{(i\mu)}\in\sym{P}_{i\mu}}\Bigr)}_{
      \mbox{dual path integral}}\,
    \Bigl(\underbrace{\prod_{(i,\mu,\nu)\in\Lambda^2}\hat
    f_{\rho_{i\mu\nu}}}_{\mbox{dual Boltzmann weight}}\Bigr)\,
    \prod_{i\in\Lambda^0}\,C(i).
\end{equation}
Here $\Irrep$ denotes a set containing one unitary representation for
each equivalence class of finite-dimensional irreducible
representations of $G$. $\sym{P}_{i\mu}$ denotes the set of all
projectors onto the different one-dimensional trivial components in
the decomposition of the tensor product~\eqref{eq_step5} into its
irreducible components. $C(i)$ describes a \emph{gauge constraint
factor} for each lattice point $i\in\Lambda^0$ which is given
by~\eqref{eq_constraint}. The coefficients $\hat f_{\rho_{i\mu\nu}}$
are defined by the character decomposition of the original Boltzmann
weight $\exp(-s(g))$,
\begin{equation}
\label{eq_dualboltzmann}
  \hat f_{\rho_{i\mu\nu}} =\dim V_{\rho_{i\mu\nu}}\,\int_G\,
  \overline{\chi^{(\rho_{i\mu\nu})}(g)}\,\exp(-s(g))\,dg.
\end{equation}
\end{theorem}

\begin{remark}
\begin{myenumerate}
\item The dual partition function can be described in words as
  follows: Colour all plaquettes with finite-dimensional irreducible
  representations of $G$ in all possible ways. Colour all links with
  projectors onto the trivial components in the tensor
  products~\eqref{eq_step5} (if there are any). The partition function
  contains a (local) dual Boltzmann weight factor which is the
  coefficient of the character expansion of the original Boltzmann
  weight for each plaquette. The partition function contains
  furthermore a (local) gauge constraint factor $C(i)$,
  see~\eqref{eq_constraint}, for each lattice point.
\item The two main differences to the Abelian case (see
  \eg~\cite{Sa80}) are the following: Firstly, in the Abelian case
  only objects on a single level, namely the plaquettes, are coloured
  with integer numbers (which characterize the finite-dimensional
  irreducible unitary representations). In the non-Abelian case we
  have to colour the plaquettes with representations and the links
  with intertwiners. The configurations of the dual model are thus
  spin foams. Note that the choice of projectors $\sym{P}_{i\mu}$
  in~\eqref{eq_step4} and~\eqref{eq_step5} agrees up to canonical
  isomorphisms with the choice of intertwiners in the definition of a
  closed spin foam as given in~\cite{Ba98}. Here the assignment of
  `incoming' and `outgoing' faces has to be made according to our
  standard orientations of plaquettes and links.

  Secondly, the integrand is not just a Boltzmann weight, but contains
  in addition the factor $C(i)$ for each lattice point. In the Abelian
  case, this factor together with the sum over projectors enforces
  co-closedness of the integer $2$-form. The dual of Abelian gauge
  theory is again a gauge theory because if this $2$-form is also
  co-exact, it can be integrated and gauge degrees of freedom appear.
  In the non-Abelian case there is no obvious integration which would
  introduce gauge degrees of freedom.
\end{myenumerate}
\end{remark}

%-------------------------------------------------------------------------------
\subsection{The dual of the non-Abelian Wilson loop}
%-------------------------------------------------------------------------------

The duality transformation of the expectation value of the non-Abelian Wilson
loop~\eqref{eq_expect} proceeds along the same lines. However, the expressions
become slightly more complicated due to the presence of the additional
integrand. In the following description of the transformation, we often refer
to the calculations for the partition function in
Section~\ref{sect_transformpart}.

The expectation value of the non-Abelian Wilson loop is given by
\begin{eqnarray}
  \bigl<\Wloop\bigr> &=& \frac{1}{Z}\,
    \Bigl(\prod_{(i,\mu)\in\Lambda^1}\int_G\,dg_{i\mu}\Bigr)\,
    \Bigl(\prod_{(i,\mu,\nu)\in\Lambda^2}f(dg_{i\mu\nu})\Bigr)\,
    \Bigl(\prod_{(j,\kappa)\in\Lambda^1}\sum_{a_{j\kappa}b_{j\kappa}}\Bigr)\nn\\
  &\times& \Bigl(\prod_{(j,\kappa)\in\Lambda^1}
        t_{a_{j\kappa}b_{j\kappa}}^{(\tau_{j\kappa})}(g_{j\kappa})\Bigr)\,
    \Bigl(\prod_{j\in\Lambda^0}
      Q^{(j)}_{(b_{j-\hat1,1}\ldots b_{j-\hat d,d}),(a_{j1}\ldots a_{jd})}\Bigr).
\end{eqnarray}
We insert the character decomposition~\eqref{eq_charexp}, employ the coproduct
and the antipode and reorganize the factors just as in the calculation for
the partition function. The result generalizes~\eqref{eq_reordering} in
which~\eqref{eq_delantipode} has been inserted:
\begin{eqnarray}
  \bigl<\Wloop\bigr> &=& \frac{1}{Z}\,
    \Bigl(\prod_{(i,\mu,\nu)\in\Lambda^2}\sum_{\rho_{i\mu\nu}\in\Irrep}\Bigr)\,
    \Bigl(\prod_{(i,\mu,\nu)\in\Lambda^2}\hat f_{\rho_{i\mu\nu}}\Bigr)\,
    \Bigl(\prod_{(j,\kappa)\in\Lambda^1}\sum_{a_{j\kappa}b_{j\kappa}}\Bigr)\\
    &&\Bigl(\prod_{(i,\mu,\nu)\in\Lambda^2}\sum_{n_{i\mu\nu},m_{i\mu\nu},p_{i\mu\nu},q_{i\mu\nu}}\Bigr)\,
    \Bigl(\prod_{j\in\Lambda^0}
        Q^{(j)}_{(b_{j-\hat1,1}\ldots b_{j-\hat d,d}),(a_{j1}\ldots a_{jd})}\Bigr)\nn\\
  &\times&\prod_{(i,\mu)\in\Lambda^2}\Biggl\{\int_G\,dg_{i\mu}\,\Biggl[
    \prod_{\lambda=1}^{\mu-1}\biggl(
      t_{m_{i-\hat\lambda,\lambda,\mu}p_{i-\hat\lambda,\lambda,\mu}}^{
         (\rho_{i-\hat\lambda,\lambda,\mu})}(g_{i\mu})\cdot
      t_{n_{i\lambda\mu}q_{i\lambda\mu}}^{(\rho_{i\lambda\mu}^\ast)}(g_{i\mu})\biggr)
      \Biggr.\Biggr.\nn\\
  &&\Biggl.\Biggl.\times\prod_{\nu=\mu+1}^d\biggl(
      t_{q_{i-\hat\nu,\mu,\nu}p_{i-\hat\nu,\mu,\nu}}^{
                 (\rho_{i-\hat\nu,\mu,\nu}^\ast)}(g_{i\mu})
      \cdot t_{n_{i\mu\nu}m_{i\mu\nu}}^{(\rho_{i\mu\nu})}(g_{i\mu})\biggr)
      \cdot\underbrace{t_{a_{i\mu}b_{i\mu}}^{(\tau_{i\mu})}(g_{i\mu})}_{
        \mbox{new factor}}
    \Biggr]\Biggr\}.\nn
\end{eqnarray}
The features which are new compared with~\eqref{eq_reordering}
and~\eqref{eq_delantipode} are the summations over $a_{j\kappa}$ and
$b_{j\kappa}$ for each link, the product over the intertwiners $Q^{(j)}$ for
each lattice point and the additional factor
$\tau_{a_{i\mu}b_{i\mu}}^{(\tau_{i\mu})}(g_{i\mu})$ in the integrand for each
link $(i,\mu)$. 

Now we can apply the projector expression for the Haar
measure~\eqref{eq_haaralg}. Compared with~\eqref{eq_step4}, the additional
factor in the integrand produces additional indices $a_{i\mu}$ and
$b_{i\mu}$ of the projectors,
\begin{eqnarray}
  &&\int_G\,dg_{i\mu}\biggl[\cdots\biggr]\nn\\
   &=& \sum_{P\in\sym{P}^\prime_{i\mu}}
     P_{(\underbrace{m_{i-\hat\lambda,\lambda,\mu}n_{i\lambda\mu}\ldots}_{
       \lambda\in\{1,\ldots,\mu-1\}})
        (\underbrace{q_{i-\hat\nu,\mu,\nu}n_{i\mu\nu}\ldots}_{
       \nu\in\{\mu+1,\ldots,d\}})\underbrace{a_{i\mu}}_{\mbox{new}}}\nn\\
   &&\quad\cdot
     P_{(\underbrace{p_{i-\hat\lambda,\lambda,\mu}q_{i\lambda\mu}\ldots}_{
       \lambda\in\{1,\ldots,\mu-1\}})
        (\underbrace{p_{i-\hat\nu,\mu,\nu}m_{i\mu\nu}\ldots}_{
       \nu\in\{\mu+1,\ldots,d\}})\underbrace{b_{i\mu}}_{\mbox{new}}}.
\end{eqnarray}
These indices correspond to the additional tensor factor in the
following decomposition: The orthogonal projectors
$P\in\sym{P}^\prime_{i\mu}$ project onto the distinct one-dimensional
trivial components in the decomposition of
\begin{equation}
\label{eq_tensorwilson}
  \underbrace{
    (\rho_{i-\hat\lambda,\lambda,\mu}\otimes\rho_{i\lambda\mu}^\ast)\otimes\cdots}_{
    \lambda\in\{1,\ldots,\mu-1\}}\otimes
  \underbrace{
    (\rho_{i-\hat\nu,\mu,\nu}^\ast\otimes\rho_{i\mu\nu})\otimes\cdots}_{
    \nu\in\{\mu+1,\ldots,d\}}\otimes\underbrace{\tau_{i\mu}}_{\mbox{new}}
\end{equation}
into its irreducible components, \cf~\eqref{eq_step5}.

Similar to the calculation for the partition function, the vector
index summations over the $n_{i\mu\nu},\ldots,q_{i\mu\nu}$ as well as
over the $a_{j\kappa}$ and $b_{j\kappa}$, the projectors
$P_{(\cdots)(\cdots)}$ and the intertwiners $Q^{(j)}$ can be
reorganized to form local expressions. This construction is entirely
analogous to the derivation of~\eqref{eq_step7}
and~\eqref{eq_constraint}. We obtain the following result which
generalizes Theorem~\ref{thm_dualpartition1}:

\begin{theorem}[Dual non-Abelian Wilson loop]
\label{thm_dualwilson1}
Let $G$ be a compact Lie group or a finite group and consider lattice
gauge theory with gauge group $G$ on a $d$-dimensional finite lattice
with periodic boundary conditions. The normalized expectation value of
the non-Abelian Wilson loop~\eqref{eq_expect} is equal to the
expression
\begin{eqnarray}
\label{eq_dualwilson1}
  \bigl<\Wloop\bigr> &=& \frac{1}{Z}\,
    \underbrace{\Bigl(\prod_{(i,\mu,\nu)\in\Lambda^2}\sum_{\rho_{i\mu\nu}\in\Irrep}\Bigr)\,
    \Bigl(\prod_{(i,\mu)\in\Lambda^1}\sum_{P^{(i\mu)}\in\sym{P}^\prime_{i\mu}}\Bigr)}_{
      \mbox{dual path integral}}\,
    \Bigl(\underbrace{\prod_{(i,\mu,\nu)\in\Lambda^2}\hat
    f_{\rho_{i\mu\nu}}}_{\mbox{dual Boltzmann weight}}\Bigr)\nn\\
  &\times&
    \prod_{i\in\Lambda^0}\,\Biggl[
    \Bigl(\prod_{\mu=1}^d\sum_{a_{i\mu}=1}^{\dim V_{\tau_{i\mu}}}
       \sum_{b_{i-\hat\mu,\mu}=1}^{\dim V_{\tau_{i-\hat\mu,\mu}}}\Bigr)\,
%    \Bigl(\prod_{(j,\kappa)\in\Lambda^1}\sum_{a_{j\kappa}b_{j\kappa}}\Bigr)\,
    Q^{(i)}_{(b_{i-\hat1,1}\ldots b_{i-\hat d,d}),(a_{i1}\ldots a_{id})}\,
    \tilde C(i)\Biggr].
\end{eqnarray}
Here $\sym{P}^\prime_{i\mu}$ denotes the set of all projectors onto
the different trivial components in the decomposition of the tensor
product~\eqref{eq_tensorwilson} into its irreducible
components. $\tilde C(i)$ describes a \emph{gauge constraint
factor} for each lattice point $i\in\Lambda^0$ which is given by
\begin{eqnarray}
\label{eq_constraint2}
  \tilde C(i) &=& \Bigl(\prod_{1\leq\mu<\nu\leq d}
           \sum_{p_{i-\hat\mu-\hat\nu,\mu,\nu}=1}^{\dim V_{\rho_{i-\hat\mu-\hat\nu,\mu,\nu}}}
           \sum_{q_{i-\hat\nu,\mu,\nu}=1}^{\dim V_{\rho_{i-\hat\nu,\mu,\nu}}}
           \sum_{m_{i-\hat\mu,\mu,\nu}=1}^{\dim V_{\rho_{i-\hat\mu,\mu,\nu}}}
           \sum_{n_{i\mu\nu}=1}^{\dim V_{\rho_{i\mu\nu}}}\Bigr)\\
  &&\prod_{\mu=1}^d
     P^{(i\mu)}_{(\underbrace{m_{i-\hat\lambda,\lambda,\mu}n_{i\lambda\mu}\ldots}_{
       \lambda\in\{1,\ldots,\mu-1\}})
        (\underbrace{q_{i-\hat\nu,\mu,\nu}n_{i\mu\nu}\ldots}_{
       \nu\in\{\mu+1,\ldots,d\}})\displaystyle a_{i\mu}}\nn\\
  &&\qquad\cdot P^{(i-\hat\mu,\mu)}_{(\underbrace{p_{i-\hat\mu-\hat\lambda,\lambda,\mu}q_{i-\hat\mu,\lambda,\mu}\ldots}_{
       \lambda\in\{1,\ldots,\mu-1\}})
        (\underbrace{p_{i-\hat\mu-\hat\nu,\mu,\nu}m_{i-\hat\mu,\mu,\nu}\ldots}_{
       \nu\in\{\mu+1,\ldots,d\}})\displaystyle b_{i-\hat\mu,\mu}}.\nn
\end{eqnarray}
\end{theorem}

\begin{remark}
The dual of the non-Abelian Wilson loop can be described in words as
follows: Just as in the Abelian case, it is not an expectation value
under the dual partition function, but looks rather like a modified
partition function. In addition to the dual partition function, there
are summations over the vector indices $a_{j\kappa}$ and $b_{j\kappa}$
which are necessary to multiply the representation matrices which form
the non-Abelian Wilson loop. The loop enters in two places. First the
intertwiners $Q^{(i)}$ appear for each lattice point
$i\in\Lambda^0$. Furthermore, the representations $\tau_{j\kappa}$ on
the links which form the non-Abelian Wilson loop, enter the tensor
product~\eqref{eq_tensorwilson}, and the corresponding indices
$a_{j\kappa}$ and $b_{j\kappa}$ thus appear
in~\eqref{eq_constraint2}. The fact that the presence of the Wilson
loop changes the constraint factors $C(i)$ is familiar from the
Abelian case. There it occurs in the expressions before the co-closed
$2$-form is integrated.
\end{remark}

%-------------------------------------------------------------------------------
\subsection{The constraints $C(i)$ on the dual lattice}
%-------------------------------------------------------------------------------

\begin{figure}[t]
\begin{center}
\input{fig/dual_2.pictex}
\mycaption{fig_dual}{%
  The relation between objects on the original lattice (solid) and on
  the dual lattice (dashed) in $d=3$. It is convenient to draw the
  points of the dual lattice shifted by half a lattice constant in all
  positive directions and to draw its axes with reversed directions.
  Then the cube dual to a point is centered around
  this point, the plaquette dual to a link is punctured by it \etc.}
\end{center}
\end{figure}

In the expression of the dual partition
function~\eqref{eq_dualpartition1} and~\eqref{eq_constraint}, the
factors $C(i)$ look very complicated. They can be understood most
easily on the dual lattice. We explain this idea for the case $d=3$
where the relevant pictures can be drawn. Analogous constructions
can be made for arbitrary $d\geq 2$.

We construct the dual lattice in the standard way which is illustrated
in Figure~\ref{fig_dual} for the case $d=3$. To each $k$-cell
$(i,\mu_1,\ldots,\mu_k)$, $1\leq\mu_1<\cdots<\mu_k\leq d$, of the
original lattice, there corresponds a $(d-k)$-cell
$(i,\nu_1,\ldots,\nu_{d-k})$, $1\leq\nu_1<\cdots<\nu_{d-k}\leq d$, of
the dual lattice such that
\begin{equation}
  \{\mu_1,\ldots,\mu_k\}\cup\{\nu_1,\ldots,\nu_{d-k}\}
  = \{1,\ldots,d\}.
\end{equation}

\begin{figure}[t]
\begin{center}
\input{fig/dualcobound_2.pictex}
\mycaption{fig_dualcobound}{%
  In $d=3$, the dual objects of the plaquettes with cobound a given
  link (solid lines) are the links in the boundary of a plaquette
  (dashed lines).}
\end{center}
\end{figure}

In the dual partition function on the original lattice, the plaquettes
are coloured with irreducible representations. The links are assigned
projectors in a certain tensor product whose factors are given by the
representations belonging to the plaquettes that cobound the
link. Conversely, on the dual lattice in $d=3$, the plaquettes
cobounding a given link correspond to the links in the boundary of a
plaquette (see Figure~\ref{fig_dualcobound}). Thus we have to colour
the links of the dual lattice with irreducible representations. The
plaquettes of the dual lattice are then assigned projectors onto the
trivial components of some tensor product. This tensor product is the
product of the representations belonging to the links in the boundary
of the plaquette.

\begin{figure}[t]
\begin{center}
\input{fig/cube_2.pictex}
\mycaption{fig_cube}{%
  The intertwiners $F\colon\rho_1\otimes\rho_2\to\rho_3\otimes\rho_4$
  between the representations $\rho_j$ associated with the links of
  the dual lattice in $d=3$. This figure indicates how the indices of
  the intertwiners $F$ are contracted in the factor $C(i)$.}
\end{center}
\end{figure}

Instead of the projectors onto trivial components, schematically
\begin{equation}
  P\colon \rho_1\otimes\rho_2\otimes\rho_3^\ast\otimes\rho_4^\ast
  \to\C
\end{equation}
we now write intertwiners
\begin{equation}
  F\colon \rho_1\otimes\rho_2\to\rho_3\otimes\rho_4,
\end{equation}
using the isomorphisms of $G$-modules $\Hom_G(V_\rho^\ast\otimes
V_\tau,\C)\cong_G\Hom_G(V_\tau,V_\rho)$. The intertwiners $F$
thus map from two links of a given plaquette to the other two links.
Note that the $F$ inherit a normalization from the $P$ coming from the
implicit inclusion
$\C\subseteq\rho_1\otimes\rho_2\otimes\rho_3^\ast\otimes\rho_4^\ast$. 

The factors $C(i)$ of~\eqref{eq_constraint} are associated with the
cubes of the dual lattice. The expression~\eqref{eq_constraint},
interpreted on the dual lattice in $d=3$, contains one intertwiner per
face of the cube as indicated in Figure~\ref{fig_cube}. In this
figure, the intertwiners $F$ are represented by double arrows
leading from two links of each plaquette to the other two links. The
arrows illustrate how the intertwiners have to be composed to account
for the contraction of the indices in~\eqref{eq_constraint}. The
remaining indices are then summed over.

A similar visualization is straight forward for the factors $\tilde
C(i)$ in~\eqref{eq_constraint2}.

% ==============================================================================
%
\section{Special cases}
%
% ==============================================================================
\label{sect_special}

%-------------------------------------------------------------------------------
\subsection{Abelian gauge theory in arbitrary dimension}
%-------------------------------------------------------------------------------

In this section we show how Theorem~\ref{thm_dualpartition1} reduces
to the well-known results for $G=U(1)$. Similar calculations are
available for $\Z$ or $\Z_n$ (Note that the transformation is
applicable to $\Z$ although $\Z$ is neither compact nor finite. This is
because $\Z$ gauge theory is dual to $U(1)$ gauge theory). 

We start with the dual partition
function~\eqref{eq_dualpartition1}. The unitary finite-dimensional
irreducible representations of $U(1)$ are all one-dimensional. They
are given by homomorphisms $g\mapsto g^k$ for $g\in U(1)$ and are
characterized by integer numbers $k\in\Z$, \ie\ $\Irrep\cong\Z$. The
dual representation is then given by $g\mapsto g^{-k}$.

Consider the tensor product~\eqref{eq_step5} and specify the
representations by integer numbers $k_{i\mu\nu}\in\Z$. Since all
irreducible representations are one-dimensional, so are their tensor
products. The question is therefore just whether or not the tensor
product~\eqref{eq_step5} is equivalent to the trivial
representation. This is the case if and only if
\begin{equation}
\label{eq_tensoru1}
  \sum_{\lambda=1}^{\mu-1}(k_{i-\hat\lambda,\lambda,\mu}-k_{i\lambda\mu})
 +\sum_{\nu=\mu+1}^d(-k_{i-\hat\nu,\mu,\nu}+k_{i\mu\nu})=0.
\end{equation}
In this case, there is exactly one projector onto a trivial component
of the tensor product which is the identity
map. If~\eqref{eq_tensoru1} does not hold, there is no such projector.

Furthermore, since all irreducible representations are
one-dimensional, the summations in the constraints $C(i)$,
see~\eqref{eq_constraint}, disappear. Moreover, the projectors
$P^{(i\mu)}$ do not have indices and are all equal to $1$ if they
exist. Thus $C(i)=1$ if~\eqref{eq_tensoru1} holds.

The partition function~\eqref{eq_dualpartition1} therefore reads for
$G=U(1)$,
\begin{eqnarray}
\label{eq_zgauge}
  Z&=&\Bigl(\prod_{(i,\mu,\nu)\in\Lambda^2}\sum_{k_{i\mu\nu}\in\Z}\Bigr)\,
    \Bigl(\prod_{(i,\mu,\nu)\in\Lambda^2}\hat f_{k_{i\mu\nu}}\Bigr)\\
  &\times&\Bigl(\prod_{(i,\mu)\in\Lambda^1}\delta\bigl(
      \sum_{\lambda=1}^{\mu-1}(k_{i-\hat\lambda,\lambda,\mu}-k_{i\lambda\mu})
      +\sum_{\nu=\mu+1}^d(-k_{i-\hat\nu,\mu,\nu}+k_{i\mu\nu})\bigr)\Bigr).\nn
\end{eqnarray}
Here we have used the notation $\delta(n)=\delta_{0,n}$ for
$n\in\Z$. 

The dual path integral thus reduces to the summation over the integer
numbers for each plaquette while the dual Boltzmann weight is again
given by the character decomposition of the original Boltzmann weight,
\begin{equation}
  \hat f_{k_{i\mu\nu}} 
  = \frac{1}{2\pi}\,\int_0^{2\pi}e^{-ik_{i\mu\nu}\phi}\exp\bigl(-s(e^{i\phi})\bigr)\,d\phi.
\end{equation}
The $\delta$-constraint ensures that the integer $2$-form
$k_{i\mu\nu}$ is co-closed. This condition provides the dual model
with the properties of a gauge theory. As is well-known, the partition
function~\eqref{eq_zgauge} describes the dual of $U(1)$ lattice gauge
theory, the so-called $\Z$ gauge theory~\cite{Sa80,PoWi91} and is here
presented on the original rather than on the dual lattice.

%-------------------------------------------------------------------------------
\subsection{Non-Abelian gauge theory in two dimensions}
%-------------------------------------------------------------------------------

In this section we demonstrate how Theorem~\ref{thm_dualpartition1}
reduces to the familiar result for non-Abelian lattice gauge theory in
$d=2$. In this case the partition function is particularly simple.

We start again with the dual partition
function~\eqref{eq_dualpartition1}. In $d=2$, there are only two
plaquettes which cobound a given link $(i,\mu)\in\Lambda^1$. Imagine
the situation of Figure~\ref{fig_linkplaqu} in $d=2$. The tensor
product~\eqref{eq_step5} therefore consists of only two factors. It
reads for links $(i,1)\in\Lambda^1$ in the $1$-direction,
\begin{equation}
\label{eq_tensord2a}
  \rho_{i-\hat 2}^\ast\otimes\rho_i,
\end{equation}
and for links $(i,2)$ in the $2$ direction,
\begin{equation}
\label{eq_tensord2b}
  \rho_{i-\hat 1}\otimes\rho_i^\ast.
\end{equation}
Here we have suppressed the last two indices of $\rho_{i\mu\nu}$ which
are always $\mu=1$ and $\nu=2$. In both
cases~\eqref{eq_tensord2a} and~\eqref{eq_tensord2b}, there are trivial
components in the tensor product if and only if
\begin{equation}
  \rho_{i-\hat 2}\cong\rho_i\qquad\mbox{resp.}\qquad
  \rho_{i-\hat 1}\cong\rho_i.
\end{equation}
Since this holds for all $i\in\Lambda^0$, the only contributions to the
partition function are given by configurations which assign the same
representation to all plaquettes. We thus have
\begin{equation}
  Z = \sum_{\rho\in\Irrep}{(\hat f_\rho)}^{|\Lambda^2|}\,
    \prod_{i\in\Lambda^0} C(i).
\end{equation}
Observe further that the projectors onto the trivial component in the
tensor products~\eqref{eq_tensord2a} and~\eqref{eq_tensord2b} are both
given by the trace, \ie\ $P^{(i\mu)}_{ab}=\frac{1}{\dim
V_\rho}\delta_{ab}$. The constraint $C(i)$ can be easily calculated:
\begin{eqnarray}
  C(i) &=& \sum_{p_{i-\hat1-\hat2}=1}^{\dim V_{\rho}}
           \sum_{q_{i-\hat2}=1}^{\dim V_{\rho}}
           \sum_{m_{i-\hat1}=1}^{\dim V_{\rho}}
           \sum_{n_{i}=1}^{\dim V_{\rho}}\,
    P^{(i,1)}_{q_{i-\hat2}n_i}\cdot P^{(i-\hat1,1)}_{p_{i-\hat1-\hat2}m_{i-\hat 1}}
    \cdot P^{(i,2)}_{m_{i-\hat1}n_i}\cdot P^{(i-\hat2,2)}_{p_{i-\hat1-\hat2}q_{i-\hat2}}\nn\\
  &=& \frac{1}{{(\dim V_\rho)}^3}.
\end{eqnarray}
Therefore the partition function reads
\begin{equation}
  Z = \sum_{\rho\in\Irrep}{(\hat f_\rho)}^{|\Lambda^2|}\,
      {(\dim V_\rho)}^{-3|\Lambda^0|}.
\end{equation}
This is the well-known result for lattice gauge theory in two
dimensions, see \eg~\cite{DrZu83}.

% ==============================================================================
%
\section{Discussion}
%
% ==============================================================================
\label{sect_outlook}

The duality transformation given in Theorems~\ref{thm_dualpartition1}
and~\ref{thm_dualwilson1} is a strong-weak duality. For example, the
character decomposition of the Boltzmann
weight~\eqref{eq_dualboltzmann} reads for the Wilson action of
$G=U(1)$, 
\begin{equation}
  \hat f_k=I_k(\beta),\qquad k\in\Z,
\end{equation}
and for the Wilson action of $G=SU(2)$ using the fundamental
representation,
\begin{equation}
  \hat f_j=2(2j+1)\,I_{2j+1}(\beta)/\beta,\qquad 2j\in\N_0.
\end{equation}
Here the representations are parameterized by integers $k$ resp.\
non-negative half-integers $j$, and $I_n(x)$ denote the modified
Bessel functions. The coefficients $\hat f_k$ resp.\ $\hat f_j$ are
positive and can thus be written $\hat f_k=\exp(-s^\ast(k))$ resp.\
$\hat f_j=\exp(-s^\ast(j))$. The $\beta$-dependence of the dual action
$s^\ast$ is such that high and low temperature regimes are exchanged
or, in the language of gauge theory, strong and weak coupling (see
\eg~\cite{Sa80,DrZu83}). However, the coupling constant does not occur
as a prefactor of the interaction terms of the dual model because its
interactions do not arise from the dual Boltzmann weight but rather
from the selection of projectors $\sym{P}_{i\mu}$ and from the factors
$C(i)$, \cf~\eqref{eq_dualpartition1}. For details about the character
decompositions of the various common actions in lattice gauge theory,
see \eg~\cite{MoMu94,DrZu83} and references therein.

Of course, it is also possible to define the action of non-Abelian
lattice gauge theory in terms of the character decomposition of its
Boltzmann weight. For example, the \emph{heat kernel action} (or
generalized Villain action) is given by the choice
\begin{equation}
  \hat f_\rho = \dim V_\rho\cdot\exp(-C_\rho/\beta),
\end{equation}
which makes the strong-weak duality manifest. Here $C_\rho$ denotes
the eigenvalue of the quad\-rat\-ic Casimir operator (in a certain
normalization) on the irreducible representation $\rho$ of $G$. Since
$C_\rho$ is essentially quadratic in the highest weight of the
representation $\rho$, it is apparent that higher representations are
exponentially suppressed in the dual path integral. The smaller
$\beta$ is chosen, the more pronounced is the suppression.  The dual
expressions~\eqref{eq_dualpartition1} and~\eqref{eq_dualwilson1} can
therefore serve as generating functions for the strong coupling
expansion. For details about strong coupling expansion techniques, see
\eg~\cite{DrZu83}.

Since the duality transformation for non-Abelian lattice gauge theory
constructed in this paper generalizes the Abelian case in the form written
with an explicit gauge constraint rather than in the form which is integrated
and exhibits gauge degrees of freedom, there is no immediate answer to the
question whether the dual model has any gauge invariance and how these degrees
of freedom could be parametrized.

A non-Abelian generalization of the integration of a closed (and
exact) $k$-cocycle to the coboundary of a $(k-1)$-cocycle up to a
gauge freedom is not easy to find. If it exists, this paper might help
to assemble information on how a non-Abelian generalization of
cohomology might look like. Interesting in this context are the ideas
developed in~\cite{GrSc98}.

Degrees of freedom in the dual model which are always present, are
related to the choice of unitary representatives $\rho\in\Rep$ of each
class of equivalent irreducible representations of $G$. The
Clebsh-Gordan coefficients which enter the analysis extensively as the
coefficients of the various projectors, depend on these choices.

In any case, the dual model can be expected to be the appropriate
starting point to search for the non-Abelian magnetic degrees of
freedom which generalize the magnetic monopoles of $U(1)$ lattice
gauge theory, and to provide a framework for a rigorous treatment of
their properties.

As far as the strong coupling expansion is concerned, the crucial
question is to what extent the limit of this expansion is compatible
with the continuum limit, \ie\ whether properties derived via the
strong coupling expansion hold (at least qualitatively) for all
couplings. Recall that the continuum limit of lattice QCD consists of
a combination of sending the inverse temperature $\beta\to\infty$ and
the lattice spacing $a\to0$.

%------------------------------------------------------------------------------
\acknowledgements
%------------------------------------------------------------------------------

Both authors are grateful to DAAD for their scholarships.
R.O.~furthermore acknowledges support by EPSRC. We would like to thank
in particular I.~Drummond, J.~Jers{\'a}k, A.~J.~Macfarlane, S.~Majid,
T.~Neuhaus and K.-G.~Schlesinger for valuable discussions, comments on
the manuscript and for drawing our attention to relevant literature.

%------------------------------------------------------------------------------

\end{document}